\documentclass[reprint,
aps,superscriptaddress,twocolumn,showpacs,nofootinbib,longbibliography]{revtex4-1}
\usepackage{graphicx,color}
\usepackage{amsmath,amsfonts,enumerate,amsthm,amssymb,bbm}
\usepackage[colorlinks=true,citecolor=blue,linkcolor=magenta]{hyperref}
\usepackage{multirow}
\usepackage{bbold}
\usepackage{braket}
\usepackage{soul}
\usepackage[caption = false]{subfig}
\usepackage{scrextend}
\usepackage{xcolor}

\mathchardef\ordinarycolon\mathcode`\:
\mathcode`\:=\string"8000
\begingroup \catcode`\:=\active
  \gdef:{\mathrel{\mathop\ordinarycolon}}
\endgroup

\theoremstyle{plain}
\newtheorem{thm}{Theorem}
\newtheorem{corol}[thm]{Corollary}

\newtheorem{propos}[thm]{Proposition}
\theoremstyle{definition}

\theoremstyle{remark}

\def\<{\langle}

\def\O{ {\cal O} }

\def\I{ \mathbb{1} }

\def\I{ \mathbbm{1} }

\def\>{\rangle}
\def\<{\langle}
\DeclareMathOperator{\Tr}{Tr} 
\DeclareMathOperator{\Var}{Var}

\renewcommand{\ket}[1]{|#1\rangle}               
\renewcommand{\bra}[1]{\langle #1|}              

\renewcommand{\vec}[1]{\boldsymbol{#1}}  

\newcommand{\new}[1]{{\color[RGB]{0,0,0}{#1}}}

\begin{document}

\title{Barren plateaus preclude learning scramblers}

\author{Zo\"{e} Holmes}
\thanks{The first three authors contributed equally to this work.}
\affiliation{Information Sciences, Los Alamos National Laboratory, Los Alamos, NM USA.}
\author{Andrew Arrasmith} 
\thanks{The first three authors contributed equally to this work.}
\affiliation{Theoretical Division, Los Alamos National Laboratory, Los Alamos, NM USA.}
\author{Bin Yan} 
\thanks{The first three authors contributed equally to this work.}
\affiliation{Center for Nonlinear Studies, Los Alamos National Laboratory, Los Alamos, NM USA.}
\affiliation{Theoretical Division, Los Alamos National Laboratory, Los Alamos, NM USA.}
\author{Patrick~J.~Coles} 
\affiliation{Theoretical Division, Los Alamos National Laboratory, Los Alamos, NM USA.}
\author{Andreas Albrecht}
\affiliation{Center for Quantum Mathematics and Physics and Department of Physics and Astronomy\\ UC Davis, One Shields Ave, Davis CA.}
\author{Andrew T. Sornborger} 
\affiliation{Information Sciences, Los Alamos National Laboratory, Los Alamos, NM USA.}

\date{\today}

\begin{abstract}
Scrambling processes, which rapidly spread entanglement through many-body quantum systems, are difficult to investigate using standard techniques, but are relevant to quantum chaos and thermalization. In this Letter, we ask if quantum machine learning (QML) could be used to investigate such processes. We prove a no-go theorem for learning an unknown scrambling process with QML, showing that \textit{any variational ansatz} is highly probable to have a barren plateau landscape, i.e., cost gradients that vanish exponentially in the system size. \new{This implies that the required resources scale exponentially even when strategies to avoid such scaling (e.g., from ansatz-based barren plateaus or No-Free-Lunch theorems) are employed.} Furthermore, we numerically and analytically extend our results to approximate scramblers. Hence, our work places generic limits on the learnability of unitaries when lacking prior information.
\end{abstract}

\maketitle

\paragraph*{Introduction.}

The growth of entanglement in many body quantum systems can rapidly distribute the information contained in the initial conditions of the system throughout a large number of degrees of freedom.
This process is known as \textit{scrambling}~\cite{HaydenPreskill, Sekino_2008, ChaosScrambler}.
In recent years scrambling has proven central not only to understanding quantum chaos but also to the study of the dynamics of quantum information~\cite{ChannelScrambling, EntanglementGrowthScrambling, DecouplingScrambling}, thermalization phenomena~\cite{ThermalizationScrambling, ThermalizationScramblingGraphene}, the black hole information paradox~\cite{HaydenPreskill, YoshidaKitaevDecode, YoshidaYaoScrambling, ScramblingIonExp}, holography~\cite{HolographyScrambler, KitaevHolog}, random circuits~\cite{ChaosByDesign,RandomMatrixScrambling,RandomCircsScrambling}, fluctuation relations~\cite{WorkStatsScrambling, JarzOTOC} and entropic uncertainty relations~\cite{EntropyUncertaintyOTOC}. However, the complexity of strongly-interacting many-body quantum systems makes scrambling rather challenging to study analytically.
Furthermore, experimental studies of scramblers are demanding given the difficulties of distinguishing scrambling from decoherence and other experimental imperfections~\cite{LandsmanIonQCScrambler, YoshidaYaoScrambling}.

\begin{figure}[t]
    \centering
    \includegraphics[width=\columnwidth]{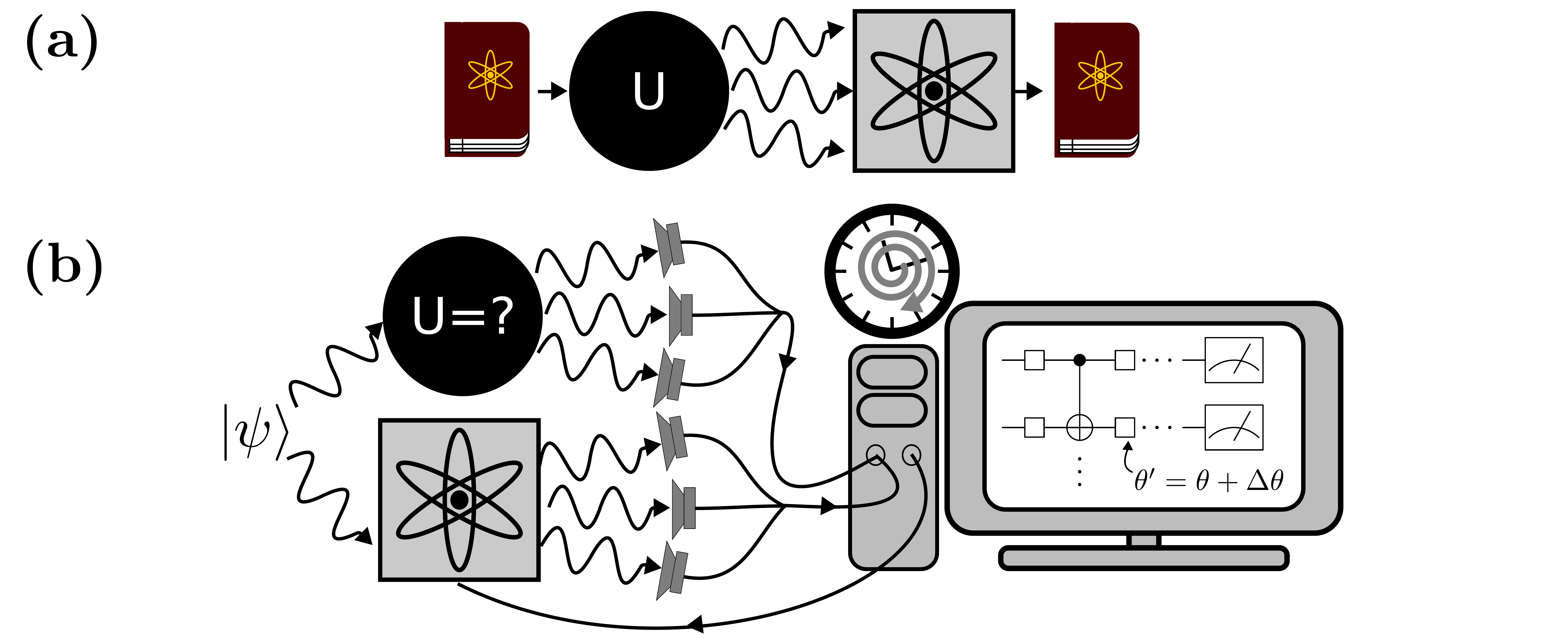}
    \caption{\textbf{Learning a scrambling unitary.} Panel (a) shows the setup of the classic Hayden-Preskill thought experiment where someone attempts to retrieve information (shown as a book) thrown into a black hole (a scrambler). If the scrambling unitary $U$ is known, then information can be retrieved. Panel (b) shows the process of attempting to learn $U$. This requires a time that is exponential in the number of quantum degrees of freedom (qubits) due to an exponentially vanishing cost gradient, see Letter below. This precludes the information retrieval shown in (a).}
    \label{fig:black_hole}
\end{figure}

Quantum computers have recently been used as a testbed for the study of scramblers~\cite{LandsmanIonQCScrambler,Yan2020, zhu2021observation, google2021information}. A possible further use of quantum computing would be to investigate scrambling using quantum machine learning (QML) methods~\cite{Preskill2018NISQ}. Here we define QML~\cite{biamonte2017quantum,BeerNN2020,poland2020no, SharmaQNFL2020} as any method that optimizes a parameterized quantum circuit by minimizing a problem-specific cost function. This includes variational quantum algorithms (VQAs) \cite{VQE,mcclean2016theory,qaoa2014,Romero,khatri2019quantum,VQSD,arrasmith2019variational,cerezo2020variationalfidelity,sharma2020noise,bravo-prieto2019,cerezo2020variational,heya2019subspace,cirstoiu2020variational,commeau2020variational,li2017efficient,endo2018variational,yuan2019theory}, which are used for numerous applications. It has recently been
shown that entanglement provides a resource to exponentially reduce the number of training states required to learn a quantum process~\cite{SharmaQNFL2020}.
Thus, one might hope that QML could prove an effective tool to study quantum scrambling. For example, Figure~\ref{fig:black_hole} shows a schematic of how this could be done for the Hayden-Preskill thought experiment~\cite{HaydenPreskill}.

However, despite the high expectations placed on QML, there remain fundamental questions concerning its scalability and breadth of applicability.
Of particular concern is the growing body of literature on the existence of \textit{barren plateaus}, i.e., regions in parameter space where cost gradients vanish exponentially as the size of the system studied increases. This phenomenon, which exponentially increases the resources required to train large scale quantum neural networks, has been demonstrated in a number of proposed architectures and classes of cost function~\cite{BarrenPlateaus2018, CerezoBP2020, SharmaBP2020,wang2020noise,cerezo2020impact,arrasmith2020effect}. 


In this paper we present a no-go theorem for the use of QML to study quantum scrambling. Namely, we show that any QML approach used to learn the unitary dynamics implemented by a typical scrambler will exhibit a barren plateau and thus be untrainable in the absence of further prior knowledge.

In contrast to the barren plateau phenomenon established in Ref.~\cite{BarrenPlateaus2018}, which is a consequence of the ansatz structure and parameter initialization strategy, our barren plateau result holds \textit{for any choice of ansatz and any initialization of parameters}. Thus, previously proposed strategies to avoid barren plateaus~\cite{Grant2019initializationBP,verdon2019learning,Volkoff2020BP}, such as correlating parameters, do not address the issue raised in our work. Our result is conceptually distinct from previous barren plateau results, and additionally provides an alternative perspective on trainability issues in QML. 

Given the close connection between chaos and randomness, our no-go theorem also applies to learning random and pseudo-random unitaries. As such, our result implies that to efficiently learn an unknown unitary process using QML, prior information about that process is required. For this reason, our work constrains the use of QML in the study of complex, arbitrary physical processes.


\bigskip

\paragraph*{Preliminaries.} 

To illustrate the machine learning task we consider, let us start by recalling the famous Hayden-Preskill thought experiment~\cite{HaydenPreskill} (Fig.~\ref{fig:black_hole}). Suppose Alice attempts to destroy a secret, encoded in a quantum state, by throwing it into Nature's fastest scrambler, a black hole. How safe is Alice's secret? Hayden and Preskill argued that if Bob knows the unitary dynamics, $U$, implemented by the black hole, and shares a maximally entangled state with the black hole, it is possible to decode Alice's secret by collecting a few additional photons emitted from the black hole. However, this prompts a second question, how might Bob learn the scrambling unitary in the first place? Here we investigate whether QML can be used to learn the scrambling unitary, $U$.

To address this, we first motivate our notion of a scrambler. A diagnostic for information scrambling that has attracted considerable recent attention is the out-of-time-ordered correlator (OTOC) \cite{Larkin1969,KitaevHolog,ChaosScrambler}, a four-point correlator with unusual time ordering,
\begin{equation}\label{eq:gen_OTOC}
     f_{\mbox{\tiny OTOC}} \equiv \langle  \tilde{X} Y  \tilde{X}^\dagger Y^\dagger \rangle \, .
\end{equation}
Here $X$ and $Y$ are local operators on different subsystems, $\tilde{X} = U^\dagger X(0) U$ is the Heisenberg evolved initial operator $X(0)$ and the average is taken over an infinite temperature state $\rho \propto \I$.

For scrambling dynamics, this quantity decays rapidly and persists at a small value. This behavior can be made more transparent by noting that if $X$ and $Y$ are both Hermitian and unitary then the OTOC can be written as
\begin{equation}
    f_{\mbox{\tiny OTOC}} = 1 - \frac{1}{2} \langle  [\tilde{X}, Y] [\tilde{X}, Y]^\dagger \rangle \, .
\end{equation}
Since $X$ and $Y$ act on different subsystems, their unevolved  commutator vanishes. However, scrambling dynamics evolve these local operators into global ones, inducing a growth of the commutator.

In this work, we are interested in learning the evolution unitary at late times, when the 
dynamics become sufficiently complex for universal structures to form. For instance, the unitary $U=\exp{(-iHt)}$ of a chaotic Hamiltonian $H$ appears increasingly random with time. Specifically, after a time scale called the scrambling time, the OTOC of a chaotic system tends to a minimal value that is equivalent to taking its average over a random distribution of unitaries~\cite{ChaosByDesign}.

The link between scrambling and randomness can be made more precise by introducing the concept of unitary designs. An ensemble of unitaries with distribution $\mu$ is a unitary $k$-design if its statistics agree with those of the Haar random distribution up to the $k$-th moment, i.e., if for any $X$,
\begin{equation}
   \int_\mu dU U^{\otimes k} (X) {U^\dagger}^{\otimes k} = \int_{Haar} dU U^{\otimes k} (X) {U^\dagger}^{\otimes k} \, .
\end{equation}
Since $f_{\mbox{\tiny OTOC}}$ in \eqref{eq:gen_OTOC} only involves the second moment of the unitary, its asymptotic smallness can be attributed to the fact that the scrambling unitary appears to be a typical element of a 2-design~\cite{ChaosByDesign}. Hence, a scrambler can be modelled as a unitary that is drawn from a distribution that forms at least a unitary 2-design~\cite{HaydenPreskill,Sekino_2008}. We remark that the dynamics of chaotic systems before reaching the scrambling time provide a model for approximate scrambling behavior~\cite{ScramblerMinimalModel}, which we study in our numerics below.


\bigskip
\paragraph*{Main results.}

Having formalized our notion of scrambling, we are now in a position to present our main result on the learnability of scramblers. The aim of QML is to minimize a problem-specific cost function that is evaluated on a quantum computer. In the context of learning unitaries, one considers an \textit{ansatz} (i.e., parameterized quantum circuit) $U(\vec{\theta})$ and a target unitary $V$. It is then natural to define the product $W(\vec{\theta}) = V^\dag U(\vec{\theta}) $, which would be proportional to the identity in the case of perfect training, i.e., when $U(\vec{\theta})$ matches the target $V$. To quantify the quality of the training, one can employ a generic cost function of the form 
\begin{equation}\label{eq:GenCost}
    C(\vec{\theta}, V) = \langle \psi | W(\vec{\theta})^{\dagger} H W(\vec{\theta}) | \psi \rangle \,,
\end{equation}
where $\ket{\psi}$ is some state and $H$ is some Hermitian operator. While in the main text we focus on the cost in~\eqref{eq:GenCost}, in Appendix~\ref{Ap:gen} we extend our results to a more general cost of the form
\new{\begin{equation}\label{eq:MoreGencost}
    C_{\rm gen}(\vec{\theta}, V) =  \sum_i p_i  \Tr[ H_i \left( \mathcal{E}_\theta \mathcal{E}_V \otimes I_R \right) \left(| \psi_i \rangle \langle \psi_i |  \right) ] \, ,
\end{equation}
where $\mathcal{E}_V(X) = V X V^\dag$ and the weighted sum with $\sum_i p_i =1$ accounts for multiple training states $\ket{\psi_i}$ and measurements $H_i$.
Here the ansatz is modeled by a general channel $\mathcal{E}_\theta$, and thus can incorporate dissipative effects such as those explored in Refs.~\cite{BeerNN2020, SharmaBP2020}.} We remark that standard cost functions for variational compiling of unitaries~\cite{khatri2019quantum,sharma2020noise} fall under the framework of our general cost in \eqref{eq:MoreGencost}, as discussed in Appendix~\ref{ap:Numericsdetails}. 

\new{Crucially, the training data $\{ \ket{\psi_i}, H_i \}$ in Eq.~\eqref{eq:MoreGencost} now acts not just on the scrambling system $S$ but can also be entangled with a reference system $R$. This is important as the No-Free-Lunch theorem introduced in Ref.~\cite{poland2020no} entails that in the absence of entanglement an exponential number of training states are required to learn an unknown unitary. Conversely, the subsequent entanglement-enhanced No-Free-Lunch theorem showed this exponential scaling can be avoided using entangled training data~\cite{SharmaQNFL2020}. In fact, a single, appropriately chosen, entangled training state may be sufficient to learn any unitary operation. This suggests it could be possible to use QML to study scramblers.}

\new{However, both the original and the entanglement-enhanced No-Free-Lunch theorems only concern the size of the training set needed to learn the action of a unitary over all states, assuming perfect training. They say nothing about the \textit{trainability} of a VQA, that is whether the cost landscape is sufficiently featured to enable the action of the unitary to be learnt on any training data. Here we address the latter.}

Barren plateaus have been proven~\cite{BarrenPlateaus2018} for a wide class of variational quantum algorithms, including those that aim to learn unitaries, whenever the training ansatz $U(\vec{\theta})$ is sufficiently random. This result is a consequence of the random nature of both the ansatz structure and the initialization of parameters, and it makes no reference to the form of a potential target unitary to be learned. Here, we prove a complementary result. Namely, if the target unitary $V$ is drawn from a sufficiently random ensemble (i.e., an ensemble that forms at least a 2-design), one also encounters a barren plateau, {\it irrespective of the training ansatz or the initialization of parameters}. Thus, regardless of the QML strategy that is employed, a typical scrambler will manifest a barren plateau. 

Suppose one wants to learn an unknown target unitary $V$ where all that is known is that it is drawn from an ensemble of scramblers $\mathbb{V}$. This corresponds to $\mathbb{V}$ forming a 2-design as noted above. Consider learning $V$ by variationally minimizing a cost $C(\vec{\theta}, V)$ of the general form in~\eqref{eq:GenCost}. The following proposition, which we prove in Appendix~\ref{Ap:VarianceGradient} of the Supplementary Material, establishes that the average gradient of the cost is zero. 
\begin{propos}\label{thm:GradientVanishes}
The average partial derivative of $C(\vec{\theta}, V)$, with respect to any parameter $\theta_k$, for an ensemble of target unitaries $\mathbb{V}$ that form a 2-design, is given by
\begin{equation}
\begin{aligned}
      \langle  \partial_{\theta_k} C(\vec{\theta}, V) \rangle_{\mathbb{V}} &=0 \, .
\end{aligned}
\end{equation}
\end{propos}

Proposition~\ref{thm:GradientVanishes} establishes that the gradient is unbiased, but this alone does not preclude the possibility of large variations in the gradient, and thus is insufficient to assess trainability. 
However, Chebyshev’s inequality bounds the probability that the partial derivative of the cost function deviates from its mean value
\begin{equation}
    P( |\partial_{\theta_k} C| \geq |x| ) \leq \frac{\Var_{\mathbb{V}}[\partial_{\theta_k} C]}{x^2}
\end{equation}
in terms of the variance of the cost partial derivative for a typical target unitary,
\begin{equation}
    \Var_{\mathbb{V}}[\partial_{\theta_k} C] = \left\langle  \left(\partial_{\theta_k} C(\vec{\theta}, V)\right)^2 \right\rangle_{\mathbb{V}} - \left\langle  \partial_{\theta_k} C(\vec{\theta}, V) \right\rangle_{\mathbb{V}}^2 \; .
\end{equation}
As a result, a vanishingly small $\Var_{\mathbb{V}}[\partial_{\theta_k} C]$ (combined with a vanishing average gradient) for all $\theta_k$ would imply that the probability that the cost partial derivative is non-zero is vanishingly small for all parameters, i.e., the cost landscape forms a barren plateau.


Indeed, this behavior is precisely what we find here. As shown in Appendix~\ref{Ap:VarianceGradient}, we prove the following.
\begin{thm}\label{thm:varienceGradient}
Consider a generic cost function $C(\vec{\theta}, V)$, Eq.~\eqref{eq:GenCost}, to learn an $n$-qubit target unitary $V$ using an arbitrary ansatz $U(\theta)$. The variance of the partial derivative of $C(\vec{\theta}, V)$, with respect to any parameter $\theta_k$, for an ensemble of target unitaries $\mathbb{V}$ that form a 2-design, is given by
\begin{equation}
\begin{aligned}
       \Var_{\mathbb{V}}[\partial_{\theta_k} C]
        = \left[\frac{2\Tr[H^2]}{2^{2n}-1}-\frac{2(\Tr[H])^2}{2^{n}(2^{2n}-1)}\right] \Var_{\chi}[-iU\partial_{\theta_k} U^\dag],
\end{aligned}\label{eq:VarGrad}
\end{equation}
where $\Var_{\mathbb{V}}$ denotes the variance over the ensemble $\mathbb{V}$, and $\Var_{\chi}$ denotes the quantum-mechanical variance with respect to the ansatz-evolved state $\ket{\chi (\vec{\theta})}=U(\vec{\theta}) \ket{\psi}$. 
\end{thm}

From Theorem~\ref{thm:varienceGradient}, we derive the following corollary on the scaling of $\Var_{\mathbb{V}}[\partial_{\theta_k} C]$. 

\begin{corol}\label{corol:varienceGradient}
Consider a generic cost function $C(\vec{\theta}, V)$, Eq.~\eqref{eq:GenCost}, to learn an $n$-qubit target unitary $V$. Without loss of generality, the ansatz can be written in the form
\begin{equation}
    U(\vec{\theta}) = \prod_{i=1}^N U_i(\vec{\theta_i}) W_i  \, ,
\end{equation}
where $\{ W_i \}$ is a chosen set of fixed unitaries and $U_i(\theta_i) = \exp{(-i\theta_i G_i)}$ with $G_i$ an Hermitian operator. If $\Tr[H^2] \in \mathcal{O}( 2^n)$ and $||G_k^2||_{\infty} \in \mathcal{O}(1)$, then 
\begin{equation}
\begin{aligned}
       \Var_{\mathbb{V}}[\partial_{\theta_k} C] \in \mathcal{O}( 2^{-n}) \, .
\end{aligned}\label{eq:ScaleGrad}
\end{equation}
\end{corol}
We note that for practical cost functions the condition $\Tr(H^2) \in \mathcal{O}( 2^n)$ holds.
Similarly, standard ans\"{a}tze use normalized generators and therefore it is reasonable to assume that $||G_k^2||_{\infty} \in \mathcal{O}(1) \,, \forall \, \, k$. 
As such, we conclude that the variance in the gradient will in general vanish exponentially with $n$, the number of qubits in the system. We note that Appendix~\ref{Ap:gen} extends this exponential scaling result to the generalized cost function in \eqref{eq:MoreGencost}. Hence, these results establish that QML approaches to learn a typical target scrambling unitary, that is a unitary drawn from a 2-design, will exhibit a barren plateau.

\begin{figure*}[t!]
    \centering
    \includegraphics[width=0.98\textwidth]{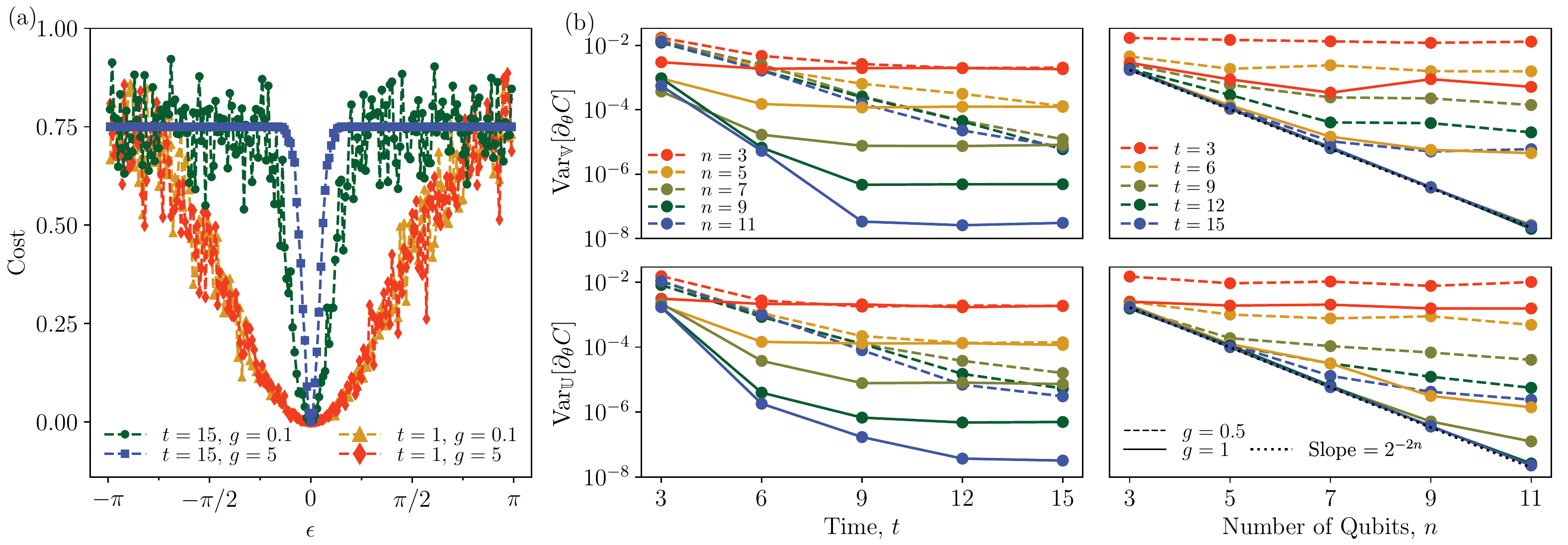}
    \caption{\textbf{Numerical Simulations of Approximate Scramblers.} a) A random cut of the landscape of the LHST cost function $C_{\rm LHST}(U, V)$ (defined in Appendix~\ref{ap:Numericsdetails}) where $V(g, t)$ is a randomly generated 9 qubit scrambler ($n=9$) modelled via Eq.~\eqref{eq:ScramblerModel} and $U(g, t)$ is an ansatz of the same form. Here $\epsilon$ is a noise parameter that determines the deviation of the ansatz parameters, $\vec{\theta}$, from the target scrambler's parameters, $\vec{\theta}^{\rm target}$. Specifically, we set $\theta_k = \theta_k^{\rm target} + \epsilon R$, where $R$ is a random number between -1 and 1. The landscape for a weak scrambler with $t=1$ and $g = 0.1$ ($g=5$) is plotted in yellow (red). Stronger scramblers with $t = 15$ and $g = 0.1$ ($g = 5$) are plotted in green (blue). b) The variance in the gradient of a single local term of the LHST cost function, $C_{\rm LHST}^{0}(U, V)$ (defined in Appendix~\ref{ap:Numericsdetails}), where $V(g, t)$ is a scrambler modelled via Eq.~\eqref{eq:ScramblerModel} with $g=0.5$ (dashed) and $g=1$ (solid) and $U(g, t)$ is an ansatz of the same form. In the first (second) row the variance is calculated over an ensemble of random target (ansatz) unitaries. In the first (second) column the variance in the gradient is plotted as a function of $t$ ($n$). As indicated by the dotted line, for large $g$ and $t$ the variance vanishes exponentially as $\Var[\partial_{\theta_k} C] \propto 2^{-2n}$.}\label{fig:VarGrad}
\end{figure*}

\medskip

\paragraph*{Numerical implementation for approximate scramblers.}
Here we extend our results by numerically studying approximate scramblers.
For concreteness, we now take a dynamical perspective and model a scrambler using a variant of the minimal model introduced in Ref.~\cite{ScramblerMinimalModel}, where the scrambling unitary 
\begin{equation} \label{eq:ScramblerModel}
    V_S(g, t) := (V_2(g) V_1 )^t 
\end{equation}
consists of alternating random gates and entangling layers. 
Specifically, the first layer is composed of a series of random single qubit rotations
\begin{equation}
    V_1 = \prod_{i} R_x^i(\theta_x^i) R_y^i(\theta_y^i) R_z^i(\theta_z^i) \, ,
\end{equation}
where $R_k^i(\theta_k^i)$ is a rotation about the $k = x, y, z$ axis of the $i_{\rm th}$ qubit and the $\{ \theta_k^i \}$ are randomly chosen angles between $0$ and $2 \pi$. The second layer is a global entangling gate 
\begin{equation}
    V_2(g) = \prod_{i<j} e^{- \frac{i g  Z_i Z_j}{\sqrt{n}} } \, ,
\end{equation}
where $Z_k$ is a Pauli operator on the $k_{\rm th}$ qubit and $n$ is the number of qubits. The degree to which $V_S(t)$ is scrambling increases with time $t$ (i.e. circuit depth), with the rate of increase determined by the entangling rate $g$. 


We consider learning a scrambler modelled by $V_S(g,t)$ using an ansatz of precisely the same structure. That is we generate a target scrambler $V_S^{\mbox{\tiny target}}$ by randomly generating a set of single qubit rotation angles $\vec{\theta}_{\mbox{\tiny target}}$ and attempt to learn the angles using an ansatz of the form $U(\vec{\theta}) = V_S(g, t)$. For concreteness, we suppose the local Hilbert-Schmidt cost~\cite{khatri2019quantum}, detailed in Appendix~\ref{ap:Numericsdetails}, is used to learn $V_S$.

In Fig.~\ref{fig:VarGrad}(a) we plot a cross-section of the cost landscape in the region around the true parameters $\vec{\theta}_{\mbox{\tiny target}}$ that minimize the cost. As we increase the degree to which the target unitary is scrambling, by increasing the duration of evolution $t$ and the strength of the entangling gates $g$, the variance in the cost visibly decreases. In the case of a highly scrambling unitary (blue) the majority of the landscape forms a \textit{barren plateau} with only a \textit{narrow gorge} where the cost dips down to its minimum. In contrast for weaker scramblers (yellow) the valley around the minimum is wider and the plateau more featured. 

In Fig.~\ref{fig:VarGrad}(b) we plot the variance of the cost partial derivative as a function of scrambling time $t$ (left) and system size $n$ (right) for varying entangling rate $g$. For completeness and for comparison, the variance is calculated both over an ensemble $\mathbb{V}$ of target unitaries (top), denoted $\Var_{\mathbb{V}}[\partial_{\theta_k} C]$, and over an ensemble $\mathbb{U}$ of random parameterized ans\"{a}tze (bottom), denoted $\Var_{\mathbb{U}}[\partial_{\theta_k} C]$. For sufficiently large $t$ and $g$ the target unitary is a perfect scrambler, and the variance of the partial derivative vanishes exponentially as $\Var_{\mathbb{V}}[\partial_{\theta_k} C] \propto 2^{-2n}$. We note that this in fact decays faster than the minimal scaling predicted by Corollary~\ref{corol:varienceGradient}, which is expected for typical scrambling ans\"{a}tze (as we discuss in Appendix~\ref{ap:ScalingAnsatzes}). Once the scrambling time is sufficient for perfect scrambling, $\Var_{\mathbb{V}}[\partial_{\theta_k} C]$ saturates. For weaker scramblers $\Var_{\mathbb{V}}[\partial_{\theta_k} C]$ similarly decreases with system size but at a slower rate. The same behavior is seen for $\Var_{\mathbb{U}}[\partial_{\theta_k} C]$, demonstrating a duality between averaging over targets and ans\"{a}tze.



In Appendix~\ref{Ap:approx}, we complement these numerical results for approximate scramblers with an analytic argument. Specifically, we show that the exponential suppression of the variance of the gradient is preserved, as long as the target ensemble is sufficiently close to a unitary 2-design.
\bigskip

\paragraph*{Discussion.}

Bob's ability to decode Alice's secret from observing relatively few emitted photons in the Hayden-Preskill thought experiment relies on the scrambling nature of the black hole's unitary dynamics. This ensures that the information contained in Alice's state is quickly distributed across the black hole. Our work establishes that the same scrambling nature inhibits Bob's ability to learn $U$ in the first place. Therefore, perhaps Alice's secret is safer than previously thought.


We find that a barren plateau will be encountered when learning a typical random (or pseudo-random) unitary with any ansatz. This no-go result provides a fundamental limit on the learnability of random processes. Related complexity theoretic arguments precluding learning black hole unitaries depend on Alice's restricted access to the full unitary \cite{Harlow2013-nl}. Our work shows that, even with complete access, the unitary is unlearnable. 

Thankfully, most physically interesting processes 
do not resemble a typical random (or pseudo-random) unitary. For example, the short time evolution of even chaotic systems form only a 1-design and consequently, as supported by our numerical results, may be learnable. Therefore the no-go theorem does not condemn QML but rather shows limits to its domain of applicability.

Crucially, Bob's situation in the Hayden-Preskill thought experiment differs from other machine learning tasks. Elsewhere, he might be able to learn more information about the target unitary, and perhaps avoid a barren plateau. When attempting to learn the dynamics of a black hole, Bob only knows that he needs to learn a scrambler and lacks a way to peek behind the event horizon to the black hole formation time, so this additional information is inaccessible. 

Methods applicable when further information is attainable remain to be studied. For example, one might try adapting the ansatz based on partial information extracted about the target. Alternatively, for certain applications, one might instead  learn a target by breaking the evolution down into shorter, more tractable, time steps. It may also be worth investigating whether barren plateaus can be avoided by only seeking partial information about the target unitary. More generally, while previous results~\cite{BarrenPlateaus2018,CerezoBP2020, SharmaBP2020} show the need to design clever ans\"{a}tze, our results highlight the need to carefully consider the properties of the target unitary.

\begin{acknowledgments}
We thank Marco Cerezo for helpful discussions. This material is based upon work supported by the U.S. Department of Energy, Office of Science, Office of High Energy Physics QuantISED program under under Contract Nos.~DE-AC52-06NA25396 and KA2401032 (ZH, AA, PJC, AA, ATS). BY acknowledges support of the U.S. Department of Energy, Office of Science, Basic Energy Sciences, Materials Sciences and Engineering Division, Condensed Matter Theory Program, and partial support from the Center for Nonlinear Studies. PJC and ATS acknowledge initial support from the Los Alamos National Laboratory (LANL) ASC Beyond Moore's Law project.

\end{acknowledgments}

\bibliography{References}

\clearpage 
\appendix
\setcounter{page}{1}
\renewcommand\thefigure{\thesection\arabic{figure}}
\setcounter{figure}{0} 

\onecolumngrid

\begin{center}
\large{ Supplementary Material for \\ ``Barren plateaus preclude learning scramblers''
}
\end{center}

\section{Preliminaries}

Here we review some prior results relevant to proving our the main results and theorems.

\medskip

\noindent \textbf{Symbolic integration.} We start by recalling formulas which allow for the symbolical integration with respect to the Haar measure on a unitary group~\cite{puchala2017symbolic}. For any $V\in \mathcal{U}(d)$ the following expressions are valid for the first two moments:  
\small
\begin{equation}\label{eq:delta}
\begin{aligned}
    \int dV \, v_{\vec{i}\vec{j}}v_{\vec{p}\vec{k}}^*&=\frac{\delta_{\vec{i}\vec{p}}\delta_{\vec{j}\vec{k}}}{d}\,,   \\
\int dV \, v_{\vec{i}_1\vec{j}_1}v_{\vec{i}_2\vec{j}_2}v_{\vec{i}_1'\vec{j}_1'}^{*}v_{\vec{i}_2'\vec{j}_2'}^{*}&=\frac{\delta_{\vec{i}_1\vec{i}_1'}\delta_{\vec{i}_2\vec{i}_2'}\delta_{\vec{j}_1\vec{j}_1'}\delta_{\vec{j}_2\vec{j}_2'}+\delta_{\vec{i}_1\vec{i}_2'}\delta_{\vec{i}_2\vec{i}_1'}\delta_{\vec{j}_1\vec{j}_2'}\delta_{\vec{j}_2\vec{j}_1'}}{d^2-1}
-\frac{\delta_{\vec{i}_1\vec{i}_1'}\delta_{\vec{i}_2\vec{i}_2'}\delta_{\vec{j}_1\vec{j}_2'}\delta_{\vec{j}_2\vec{j}_1'}+\delta_{\vec{i}_1\vec{i}_2'}\delta_{\vec{i}_2\vec{i}_1'}\delta_{\vec{j}_1\vec{j}_1'}\delta_{\vec{j}_2\vec{j}_2'}}{d(d^2-1)}\,,
\end{aligned}
\end{equation}
\normalsize
where $v_{\vec{i}\vec{j}}$ are the matrix elements of $V$. Assuming $d=2^n$, we use the notation $\vec{i} = (i_1, \dots i_n)$ to denote a bitstring of length $n$ such that $i_1,i_2,\dotsc,i_{n}\in\{0,1\}$. 

\medskip

\noindent \textbf{Useful Identities.} 
We use the following identities, which can be derived using Eq.~\eqref{eq:delta} (see~\cite{CerezoBP2020, Yan2020-ov} for a review):
\small
\begin{align}
&\int dV \Tr[V A V^\dag B] = \frac{\Tr[A] \Tr[B]}{d} \label{eq:Id1} \\
&\int d V \operatorname{Tr}\left[V A V^{\dagger} B\right] \operatorname{Tr}\left[V C V^{\dagger} D\right] = \frac{\operatorname{Tr}[A] \operatorname{Tr}[B] \operatorname{Tr}[C] \operatorname{Tr}[D]+\operatorname{Tr}[A C] + \operatorname{Tr}[B D]}{d^{2}-1}-\frac{\operatorname{\Tr}[A C] \operatorname{Tr}[B] \operatorname{Tr}[D]+\operatorname{Tr}[A] \operatorname{Tr}[C] \operatorname{Tr}[B D]}{d\left(d^{2}-1\right)} \label{eq:Id2} \\
&\int dV (V\otimes \I) A (V^\dag\otimes \I) \, B = \frac{\I_S \otimes \text{Tr}_S[A]}{d_S} B \, . \label{eq:SubspaceId1}
\end{align}
\normalsize
where $A, B, C$, and $D$ are linear operators on a $d$-dimensional Hilbert space, $d = 2^n$ is the dimension of $V$, and $\text{Tr}_S$ is a partial trace over the $d_S = 2^n$ dimensional system $S$.

\new{Let $A, B, C, D \in \mathcal{L}(\mathcal{H})$, where $\mathcal{H}$ is a $d^2$-dimensional Hilbert space. 
Then from Eq.~\eqref{eq:delta}, we derive the following integral:
\begin{equation}\label{eq:subspaceId2}
\begin{aligned}
    I =& \int dV \Tr[(V\otimes\I) A (V^\dagger \otimes \I)B]\Tr[(V\otimes\I) C (V^\dagger \otimes \I)D]\\
    =&\frac{1}{d^2-1} \Big(\Tr\left[\Tr_{S_1S_2}[A\otimes C]\Tr_{S_1S_2}[B\otimes D]\right]+\Tr\left[\Tr_{S_1S_2}[(A\otimes C)P]\Tr_{S_1S_2}[(B\otimes D)P]\right] \Big) \\
    &-\frac{1}{d(d^2-1)}\Big(\Tr\left[\Tr_{S_1S_2}[(A\otimes C)P]\Tr_{S_1S_2}[B\otimes D]\right]+\Tr\left[\Tr_{S_1S_2}[A\otimes C]\Tr_{S_1S_2}[(B\otimes D)P]\right] \Big). 
\end{aligned}
\end{equation}
where $P$ is the subsystem swap operator $  P\ket{i}\ket{j}\ket{k}\ket{l}=\ket{k}\ket{j}\ket{i}\ket{l}$. 

\begin{proof}
We begin by writing
\begin{equation}
\begin{aligned}
    \mathcal{I}&=\int dV \Tr[(V\otimes\I) A (V^\dagger \otimes \I)B]\Tr[(V\otimes\I) C (V^\dagger \otimes \I)D]\\
    &=\sum_{i_1 i'_1 j_1 j_1'}\sum_{i_2 i'_2 j_2 j_2'}\int dV V_{i_1 j_1}V^*_{i'_1 j'_1}V_{i_2 j_2}V^*_{i'_2 j'_2}\\
    &\qquad\qquad\qquad\qquad\quad\cdot\Tr[(\ket{i_1}\bra{j_1}\otimes\I) A (\ket{j'_1}\bra{i'_1} \otimes \I)B]\Tr[(\ket{i_2}\bra{j_2}\otimes\I) C (\ket{j'_2}\bra{i'_2} \otimes \I)D] \, .
\end{aligned}
\end{equation}
In the last line here we are expanding the unitaries in the computational basis. Then using Eq.~\eqref{eq:delta} to evaluate the integral, we have:
\begin{equation}
\begin{aligned}
    \mathcal{I}=&\frac{1}{d^2-1} \sum_{i_1 i_2 j_1 j_s}\left(\Tr[(\ket{i_1}\bra{j_1}\otimes\I) A (\ket{j_1}\bra{i_1} \otimes \I)B]\Tr[(\ket{i_2}\bra{j_2}\otimes\I) C (\ket{j_2}\bra{i_2} \otimes \I)D]\right.\\
    &\left.\qquad\qquad\quad+\Tr[(\ket{i_1}\bra{j_1}\otimes\I) A (\ket{j_2}\bra{i_2} \otimes \I)B]\Tr[(\ket{i_2}\bra{j_2}\otimes\I) C (\ket{j_1}\bra{i_1} \otimes \I)D] \right) \\
    &-\frac{1}{d(d^2-1)} \sum_{i_1 i_2 j_1 j_s}\left(\Tr[(\ket{i_1}\bra{j_1}\otimes\I) A (\ket{j_2}\bra{i_1} \otimes \I)B]\Tr[(\ket{i_2}\bra{j_2}\otimes\I) C (\ket{j_1}\bra{i_2} \otimes \I)D]\right.\\
    &\left.\qquad\qquad\qquad\qquad+\Tr[(\ket{i_1}\bra{j_1}\otimes\I) A (\ket{j_1}\bra{i_2} \otimes \I)B]\Tr[(\ket{i_2}\bra{j_2}\otimes\I) C (\ket{j_2}\bra{i_1} \otimes \I)D]\right)\\
    =&\frac{1}{d^2-1} \sum_{i_1 i_2 j_1 j_s}\left(\Tr[(\ket{i_1}\bra{j_1}\otimes\I\otimes\ket{i_2}\bra{j_2}\otimes\I) (A\otimes C) (\ket{j_1}\bra{i_1} \otimes \I \otimes\ket{j_2}\bra{i_2} \otimes \I)(B\otimes D)]\right.\\
    &\left.\qquad\qquad\quad+\Tr[(\ket{i_1}\bra{j_1}\otimes\I\otimes\ket{i_2}\bra{j_2}\otimes\I) (A\otimes C) (\ket{j_2}\bra{i_2} \otimes \I \otimes\ket{j_1}\bra{i_1} \otimes \I)(B\otimes D)] \right) \\
    &-\frac{1}{d(d^2-1)} \sum_{i_1 i_2 j_1 j_s}\left(\Tr[(\ket{i_1}\bra{j_1}\otimes\I\otimes \ket{i_2}\bra{j_2}\otimes\I) (A\otimes C) (\ket{j_2}\bra{i_1} \otimes \I \otimes \ket{j_1}\bra{i_2} \otimes \I )(B\otimes D)]\right.\\
    &\left.\qquad\qquad\qquad\qquad+\Tr[(\ket{i_1}\bra{j_1}\otimes\I\otimes \ket{i_2}\bra{j_2}\otimes\I) (A\otimes C) (\ket{j_1}\bra{i_2} \otimes \I\otimes \ket{j_2}\bra{i_1} \otimes \I)(B\otimes D)]\right) \, ,\\
\end{aligned}
\end{equation}
which is equivalent to the right hand side of Eq.~\eqref{eq:subspaceId2} as claimed. \end{proof}}

\section{Proofs of results presented in the main text}\label{Ap:VarianceGradient}

\subsection{Proof of Proposition~\ref{thm:GradientVanishes}}

\begin{proof}
We start by switching the order of the average and the derivative,
\begin{equation}\label{eq:GradientVanishesProof}
\begin{aligned}
         \langle  \partial_{\theta_k} C(\vec{\theta}, V) \rangle_{\mathbb{V}} &= \int dV \partial_{\theta_k} C(\vec{\theta}, V)  \\ &= \partial_{\theta_k} \int dV \Tr \left[U^\dag(\vec{\theta}) V H V^\dag U(\vec{\theta}) \rho \right],
\end{aligned}
\end{equation}
where $\rho=|\psi\rangle\langle\psi|$. The Haar integral over $V$ can be evaluated using Eq.~\eqref{eq:Id1}. Thus we are left with 
\begin{equation}\label{eq:GradientVanishes}
\begin{aligned}
        \langle  \partial_{\theta_k} C(\vec{\theta}, V) \rangle_{\mathbb{V}}
        &=    \frac{\text{Tr}[H]}{2^n} \partial_{\theta_k} \Tr\left[\rho\right]\\
        &=0 \, ,
\end{aligned}
\end{equation}
as claimed. 
\end{proof}

\subsection{Proof of Theorem~\ref{thm:varienceGradient}}\label{app:varienceGradient}


\begin{proof}
The variance of the cost function 
\begin{equation}\label{app:cost}
    C(\vec{\theta}, V) = \langle \psi | U^\dag(\vec{\theta}) V H V^\dag U(\vec{\theta}) | \psi \rangle \,
\end{equation}
is given by
\begin{equation}
\begin{aligned}
        \Var_{\mathbb{V}}[\partial_{\theta_k} C] &\equiv \int dV\  (\partial_{\theta_k} C)^2 \\
        & = \int  dV\ \text{Tr}\left[  \frac{\partial  U^\dag(\vec{\theta}) V H V^\dag U(\vec{\theta})\rho}{\partial \theta_k}   \right]^2, 
\end{aligned}
\end{equation}
where $\rho = |\psi\rangle\langle\psi|$. Note that we have used the result of Proposition~\ref{thm:GradientVanishes} to neglect the squared mean term.
This integral can be evaluated using Eq.~\eqref{eq:Id2}
to obtain 
\begin{equation}\label{app:variance}
\begin{aligned}
       \Var_{\mathbb{V}}[\partial_{\theta_k} C]
        = \left[\frac{2\text{Tr}[H^2]}{2^{2n}-1}-\frac{2\text{Tr}[H]^2}{2^n(2^{2n}-1)}\right]\Var_{\chi}[-iU\partial_{\theta_k} U^\dag],
\end{aligned}
\end{equation}
where the variance on the RHS is evaluated with respect to the state $\ket{\chi} = U \ket{\psi}$.
\end{proof}


\subsection{Proof of Corollary~\ref{corol:varienceGradient}}\label{app:corol}

\begin{proof}
The term $\Var_{\chi}[-iU\partial_{\theta_k} U^\dag]$ can be evaluated by considering a layered, parameterized circuit structure of the form 
\begin{equation}
    U(\vec{\theta}) = \prod_{i=1}^N U_i(\vec{\theta_i}) W_i  \, .
\end{equation}
Here $\{ W_i \}$ is a chosen set of fixed unitaries and $U_i(\theta_i) = \exp{(-i\theta_i G_i)}$ where $G_i$ is a Hermitian operator. 
Next consider a bipartite cut made at the $k_{\rm th}$ layer of this circuit structure and write 
\begin{equation}\label{eq:structure}
    U(\vec{\theta}) \equiv U_{L}^k(\vec{\theta}) U_R^k(\vec{\theta}) 
\end{equation}
where 
\begin{equation}\label{eq:structureSplit}
\begin{aligned}
 &U_L^k(\vec{\theta}) = \prod_{i=k+1}^N U_i(\theta_i) W_i \ \ \text{and} \ \ U_R^k(\vec{\theta}) = \prod_{i=1}^k U_i(\theta_i) W_i \, .
\end{aligned}
\end{equation}
The term $\Var_{\chi}[-iU\partial_{\theta_k} U^\dag]$ evaluates to
\begin{equation}
    \Var_{\chi}[-iU\partial_{\theta_k} U^\dag] = \Var_{\chi^k_R}[G_k]
\end{equation}
where $\Var_{\chi^k_R}[G_k]$ is the variance of $G_k$ with respect to the state $\ket{\chi_R^k} = U_R^k |\psi \rangle $. Thus we have that the variance in the cost is given by 
\begin{equation}
\begin{aligned}
        \Var_{\mathbb{V}}[\partial_{\theta_k} C] = \left[\frac{2\text{Tr}[H^2]}{2^{2n}-1}-\frac{2(\text{Tr}[H])^2}{2^n(2^{2n}-1)}\right]\Var_{\chi^k_R}[G_k] \, .
\end{aligned}
\end{equation}
We further note that $\Var_{\chi^k_R}[G_k]$ can be bounded as follows
\begin{equation}
\begin{aligned}
\Var_{\chi^k_R}[G_k] = \bra{\chi^k_R} {G_k}^2 \ket{\chi^k_R} - \bra{\chi^k_R} {G^k} \ket{\chi^k_R}^2 \leq \bra{\chi^k_R} {G_k}^2 \ket{\chi^k_R} \leq ||G_k^2||_{\infty} \, 
\end{aligned}
\end{equation}
where $||X||_{\infty}$ denotes the infinity norm of $X$, i.e. its largest eigenvalue. 
Therefore if $||G_k^2||_{\infty} \in \mathcal{O}(1)$, it follows that $\Var_{\chi^k_R}[G_k] \in \mathcal{O}(1)$. Assuming additionally that $\Tr[H^2] \in \mathcal{O}(2^n)$, from which it follows that $\Tr[H] \in \mathcal{O}(2^n)$, gives 
\begin{equation}
\begin{aligned}
        \Var_{\mathbb{V}}[\partial_{\theta_k} C] \in \O(2^{-n})  \, ,
\end{aligned}
\end{equation}
as claimed. 
\end{proof}


\section{Extension of main results to the generalized cost $C_{\rm gen}$}\label{Ap:gen}

\new{In this section we extend our results on the exponential suppression for the cost function gradient to a generalized cost function of the form,
\begin{equation}\label{app:gencost_chan}
    C_{\rm gen}(\vec{\theta}, V) =  \sum_i p_i  \Tr[ H_i \left( \mathcal{E}_\theta \mathcal{E}_V \otimes I_R \right) \left(| \psi_i \rangle \langle \psi_i |  \right) ] \, ,
\end{equation}
were $\mathcal{E}_\theta$ is a channel implemented by the ansatz and $\mathcal{E}_V(X) = V^\dag X V$ is the channel capturing evolution under the inverse of the scrambling unitary.
Since the ansatz is now represented by a channel it can incorporate non-unitary effects such as dissipation. The weighted sum with $\sum_i p_i =1$ accounts for multiple training states $\ket{\psi_i}$ and measurements $H_i$.
Crucially, the training data $\{ \ket{\psi_i}, H_i \}$ now acts not just on the scrambling system $S$ but can also be entangled with a reference system $R$.
Such entangled training data was proven beneficial for quantum machine learning in Ref.~\cite{SharmaQNFL2020} and the power of dissipative quantum neural networks has been explored in Refs~\cite{BeerNN2020, SharmaBP2020}.}

\medskip

\new{It follows from Stinespring's dilation theorem that the ansatz channel can be modelled as arising from a unitary evolution on a larger (dilated) system $D$, 
\begin{equation}\label{eq:ChannelAnsatz}
\mathcal{E}_\theta( X ) := \Tr_D[U(\theta) (X \otimes |0 \rangle \langle 0 |_D )  U(\theta)^\dagger]   \, .
\end{equation}
As such the generalized cost can alternatively be written as 
\begin{equation}\label{app:gencost}
    C_{\rm gen}(\vec{\theta}, V) =  \sum_i p_i  \langle \Psi^{(i)} | ((U^\dag(\vec{\theta}) ( V \otimes \I_D))\otimes \I_R) (H_i\otimes \I_D) ((V^\dag \otimes \I_D )  U(\vec{\theta}))\otimes \I_R )| \Psi^{(i)} \rangle \, .
\end{equation}
Here we have split the statespace into three systems: the $2^n$ dimensional system $S$ that the target unitary $V$ acts on, a system $D$ used to model dissipation, and a reference system $R$. The measurement operators $\{H_i\}$ act non-trivially on $S\otimes R$, the ansatz $U^\dag(\vec{\theta})$ acts on $S\otimes D$, and the states $\{ \ket{\Psi_i} \}$ span the full space $S \otimes D\otimes R$. We will work with the cost in this form to prove our main results.} 

\medskip

In analogy to results shown in the main text, we have proven the following propositions and theorems on the cost landscape of $C_{\rm gen}(\vec{\theta}, V) $. 

\begin{propos}\label{thm:GenCostGradientVanishes}
The average partial derivative of the general cost $C_{\rm gen}(\vec{\theta}, V)$, with respect to any parameter $\theta_k$, for an ensemble of target unitaries $\mathbb{V}$ that form a 2-design, is given by
\begin{equation}
\begin{aligned}
      \langle  \partial_{\theta_k} C_{\rm gen} (\vec{\theta}, V) \rangle_{\mathbb{V}} &=0 \, .
\end{aligned}
\end{equation}
\end{propos}

\begin{thm}\label{thm:ScalingGenCost}
Consider the general cost function $C_{\rm gen}(\vec{\theta}, V)$ to learn an $n$-qubit target unitary $V$. Without loss of generality, the ansatz can be written in the form \new{of Eq.~\eqref{eq:ChannelAnsatz} where} $ U(\vec{\theta}) = \prod_{i=1}^N U_i(\vec{\theta_i}) W_i $ with $\{ W_i \}$ is a chosen set of fixed unitaries and $U_i(\theta_i) = \exp{(-i\theta_i G_i)}$ with $G_i$ an Hermitian operator. Let us decompose the measurement operators as $H_i = \sum_j q_j H^{S}_{ij} \otimes H^R_{ij},\ \sum_j q_j = 1$, where $S$ and $R$ label the corresponding Hilbert subspace in the tensor product $U^\dag V\otimes \I$. Let $w_{ijk}$ denote the eigenvalues of $H_{ij}^{R}$.
If $\Tr[H_i^{S} H_j^{S}  ] \in \mathcal{O}( 2^n)$, $\Tr[H_i^{S}] \in \mathcal{O}(2^n)$, $w_{ij} \in \mathcal{O}(1) $ and $||G_k^2||_{\infty} \in \mathcal{O}(1)$, then 
\begin{equation}
\begin{aligned}
       \Var_{\mathbb{V}}[\partial_{\theta_k} C_{\rm gen}] \in \mathcal{O}( 2^{-n}) \, .
\end{aligned}\label{eq:ScaleGradApp}
\end{equation}
\end{thm}

\subsection{Proof of Proposition~\ref{thm:GenCostGradientVanishes}}

\begin{proof}
Switching the order of the average and the derivative, the averaged gradient of Eq.~\eqref{app:gencost} reads,
\new{\begin{equation}
\begin{aligned}
         \langle  \partial_{\theta_k} C_{\rm gen}(\vec{\theta}, V) \rangle_{\mathbb{V}} &=
         \int dV \partial_{\theta_k} C_{\rm gen}(\vec{\theta}, V)  \\ &= \partial_{\theta_k} \Tr \left[ \sum_i p_i \int dV  (U^\dag(\vec{\theta})(V \otimes \I_D)\otimes \I_R) (H_i\otimes \I_D) ((V^\dag  \otimes \I_D )   U(\vec{\theta})\otimes \I_R) \rho_i    \right],
\end{aligned}
\end{equation}}
where $\rho_i = |\psi_i\rangle\langle\psi_i|$. Each integral term can be performed using Eq.~\eqref{eq:SubspaceId1}. Thus
\new{\begin{equation}
\begin{aligned}
    \langle  \partial_{\theta_k} C_{\rm gen}(\vec{\theta}, V) \rangle_{\mathbb{V}} &= \partial_{\theta_k} \Tr \left[ \sum_i p_i \int dV  (U^\dag(\vec{\theta})(V \otimes \I_D)\otimes \I_R) (H_i\otimes \I_D) ((V^\dag  \otimes \I_D )   U(\vec{\theta})\otimes \I_R) \rho_i    \right]\\
   &= \partial_{\theta_k} \Tr \left[ \sum_i p_i \frac{\I_S\otimes \I_D \otimes \text{Tr}_S[ H_i]}{2^n} ((U^\dag(\vec{\theta})  U(\vec{\theta}))\otimes I_R) \rho_i \right] \,  \\
    &= \partial_{\theta_k} \Tr \left[ \sum_i p_i \frac{\I_S\otimes \I_D \otimes \text{Tr}_S[ H_i]}{2^n} \rho_i \right] = 0 \, .
\end{aligned}
\end{equation}}

\end{proof}



\subsection{Proof of Theorem~\ref{thm:ScalingGenCost}}

\begin{proof}
\new{Without loss of generality, each measurement operator $H_i$ can be decomposed as
\begin{equation}\label{eq:Decomp}
  H_i = \sum_j q_{ij} H^S_{ij} \otimes H^R_{ij},\ \sum_j q_{ij} = 1 \, ,
\end{equation}
into operators acting on the system $S$ and reference $R$ respectively. With this decomposition, the cost \eqref{app:gencost} can be recast into the form
 \begin{equation}\label{app:gencost2}
\begin{aligned}
      C_{\rm gen}(\vec{\theta}, V) &=  \sum_{ij} p_i q_{ij} \langle \psi_{ij} |\new{ (U^\dag(\vec{\theta})} \otimes \I_R)(  V \otimes \I_D\otimes \I_R) (H^{S}_{ij} \otimes \I_D \otimes H^R_{ij}) (V^\dag \otimes \I_D \otimes \I_R  ) (\new{U(\vec{\theta})} \otimes \I_R)| \psi_{ij} \rangle
\end{aligned}
 \end{equation}
where we define $\ket{\psi_{ij}} = \ket{\psi_i}$ for all $j$.
 
Now, for each index $i,j$, we can further decompose the corresponding input states as
 \begin{equation}
     |\psi_{ij}\rangle = \sum_k \alpha_{ijk} |\psi^{SD}_{ijk}\rangle \otimes |\psi^R_{ijk}\rangle, \ \sum_k |\alpha_{ijk}|^2 = 1.
 \end{equation}
 Here, for each pair of $\{i,j\}$, we also have the freedom to choose $\{|\psi^R_{ijk}\rangle\}_k$ as a set of orthogonal eigenstates of $H^R_{ij}$. Note that this may not be a Schmidt decomposition and $\{|\psi^{SD}_{ijk}\rangle\}_k$ in general are not orthogonal. This allows us to fully factorize cost
 \eqref{app:gencost2}, i.e.,
  \begin{equation}
\begin{aligned}
      C_{\rm gen}(\vec{\theta}, V) 
      = \sum_{ijk} p_i q_{ij} |\alpha_{ijk}|^2  \langle \psi^{SD}_{ijk} | U^\dag(\vec{\theta}) ((V H^S_{ij} V^\dag) \otimes I_D) U(\vec{\theta})| \psi^{SD}_{ijk} \rangle \langle \psi^R_{ijk} |H^R_{ij}|\psi^R_{ijk} \rangle \, .
\end{aligned}
 \end{equation}
Since $\{|\psi^R_{ijk}\rangle\}_k$ is chosen as a set of the eigenstates of $H^R_{ij}$, $ \langle \psi^R_{ijk} |H^R_{ij}|\psi^R_{ijk} \rangle \equiv w_{ijk}$ are the eigenvalues of $H^R_{ij}$. We now relabel the triple of $\{i,j,k\}$ as $\{l\}$, and define $w_l = w_{ijk}$, $H^S_{l} = H^S_{ijk}=H^S_{ij}$ for all $k$ and $\tilde{p}_l = p_i q_{ij} |\alpha_{ijk}|^2$, with $\sum_l \tilde{p}_l =1$. On doing so, the generalized cost takes the form
\begin{equation}
    C_{\rm gen}(\vec{\theta}, V) =  \sum_{l} \tilde{p}_l w_l \langle \psi^{SD}_l | U^\dag(\vec{\theta})( (V H^S_l  V^\dag )\otimes I_D) U(\vec{\theta} )| \psi^{SD}_l \rangle  \,. 
\end{equation}

The gradient of this cost reads
\begin{equation}\label{app:gengrad}
    \partial_{
    \theta_k} C_{\rm gen}(\vec{\theta}, V) = \sum_i \tilde{p}_l w_l \text{Tr} \left[ ( (V H^S_l  V^\dag )\otimes I_D) \frac{\partial U(\vec{\theta}) \rho_l^{SD} U^\dag(\vec{\theta})}{\partial \theta_k } \right],
\end{equation}
where $\rho_l^{SD} = |\psi_l^{SD}\rangle\langle\psi_l^{SD}|$. This
gives the form of its variance 
\begin{equation}
\begin{aligned}
    \Var_{\mathbb{V}}[\partial_{\theta_k} C_{\rm gen}(\vec{\theta}, V)] &= \int dV\  (\partial_{\theta_k} C_{\rm gen})^2 \\
    &= \sum_{lm}  \tilde{p}_l w_l  \tilde{p}_m w_m \int dV\ \text{Tr} \left[ (VH_l^SV^\dag \otimes I_D) \frac{\partial U(\vec{\theta}) \rho_l^{SD}  U^\dag(\vec{\theta})}{\partial \theta_k } \right] \text{Tr} \left[ (VH_m^S V^\dag  \otimes I_D) \frac{\partial U(\vec{\theta}) \rho_m^{SD}  U^\dag(\vec{\theta})}{\partial \theta_k } \right] \\
    &= \sum_{lm} \tilde{p}_l w_l  \tilde{p}_m w_m S_{lm}^k \, .
\end{aligned}
\end{equation}
Here we have used the result of Proposition~\ref{thm:GenCostGradientVanishes} to neglect the squared mean term. 

Using the identity presented in Eq.~\eqref{eq:subspaceId2}, each integral term  can be evaluated similarly as}
\begin{equation}
\begin{aligned}
    S_{lm}^k=\left[\frac{2\text{Tr}[H_l^S H_m^S]}{2^{2n}-1}-\frac{2\text{Tr}[H_l^S]\text{Tr}[H_m^S]}{2^n(2^{2n}-1)}\right]\left[\text{Tr}\left[(-iU\partial U^\dag)^2\chi_{lm}^{SD}\right] \text{Tr}\left[\chi_{lm}^{SD}\right] - (\text{Tr}\left[-iU\partial U^\dag\chi_{lm}^{SD}\right])^2  \right], \\
\end{aligned}
\end{equation}
where $\chi_{lm}^{SD}=U|\psi_l^{SD}\rangle\langle \psi_m^{SD}|U^{\dag}$. Denote $J_k=-iU\partial_{\theta_k} U^{\dag}$. The second factor in the above solution can be written in a compact form  
\begin{equation}
    \Var_{\chi_{lm}^{SD}}[J_k] \equiv \text{Tr}\left[\chi_{lm}^{SD} J_k^2\right]\text{Tr}\left[\chi_{lm}^{SD}\right] - \left(\text{Tr}\left[\chi_{lm}^{SD} J_k\right]\right)^2 ,
\end{equation}
the variance of $J_k$ with respect to $\chi_{lm}^{SD}$. It can be seen that for the terms with $l=m$, it reduces to expression \eqref{app:variance}.
Thus, the variance of the gradient takes the form 
\begin{equation}\label{eq:full_gen_variance}
     \Var_{\mathbb{V}}[\partial_{\theta_k} C_{\rm gen}(\vec{\theta}, V)] =  \sum_{lm} \tilde{p}_l w_l \tilde{p}_m  w_m \left[\frac{2\text{Tr}\left[H_l^S H_m^S\right]}{2^{2n}-1}-\frac{2\text{Tr}\left[H_l^S\right] \text{Tr}\left[H_m^S\right] }{2^n(2^{2n}-1)}\right] \Var_{\chi_{lm}^{SD}}[J_k] \, .
\end{equation}

The proof is completed by evaluating the asymptotic scaling of Eq.~\eqref{eq:full_gen_variance}. We begin by noting that
\begin{equation}
\begin{aligned}
     \Var_{\mathbb{V}}[\partial_{\theta_k} C_{\rm gen}(\vec{\theta}, V)] &= \left| \sum_{lm} \tilde{p}_l w_l \tilde{p}_m  w_m \left[\frac{2\text{Tr}\left[H_l^S H_m^S\right]}{2^{2n}-1}-\frac{2\text{Tr}\left[H_l^S\right] \text{Tr}\left[H_m^S\right] }{2^n(2^{2n}-1)}\right] \Var_{\chi_{lm}^{SD}}[J_k] \right|\\
     &\leq  \sum_{lm} \tilde{p}_l \tilde{p}_m |w_l|  |w_m| \left|\left[\frac{2\text{Tr}\left[H_l^S H_m^S\right]}{2^{2n}-1}-\frac{2\text{Tr}\left[H_l^S\right] \text{Tr}\left[H_m^S\right] }{2^n(2^{2n}-1)}\right]\right| \Var_{\chi_{lm}^{SD}}[J_k] \\
     &\leq  \sum_{lm} \tilde{p}_l \tilde{p}_m |w_l|  |w_m| \left[\frac{2\left|\text{Tr}\left[H_l^S H_m^S\right]\right|}{2^{2n}-1}+\frac{2\left|\text{Tr}\left[H_l^S\right] \right|\left|\text{Tr}\left[H_m^S\right] \right|}{2^n(2^{2n}-1)}\right] \Var_{\chi_{lm}^{SD}}[J_k]\, .
\end{aligned}
\end{equation}
Then, defining $w_{\text{max}}=\text{max}\{|w_k|\}$,  $X_{\rm max} =\text{max}\{ |\text{Tr}\left[H_l^S H_m^S\right] |\}$, and $ Y_{\max} = \text{max} \{|\text{Tr}\left[H_l^S\right]| |\text{Tr}\left[H_m^S\right]|\}$, we have that 
\begin{equation}
\begin{aligned}
     \Var_{\mathbb{V}}[\partial_{\theta_k} C_{\rm gen}(\vec{\theta}, V)] &\leq w_{\text{max}}^2 \left[\frac{2 X_{\max}}{2^{2n}-1}+\frac{2 Y_{\max}}{2^n(2^{2n}-1)}\right]\sum_{lm} \tilde{p}_l \tilde{p}_m   \Var_{\chi_{lm}^{SD}}[J_k]\, .
\end{aligned}
\end{equation}
As before, we consider a layered, parameterized circuit structure of the form $ U(\vec{\theta}) = \prod_{i=1}^N U_i(\vec{\theta_i}) W_i $ where $\{ W_i \}$ is a chosen set of fixed unitaries, $U_i(\theta_i) = \exp{(-i\theta_i G_i)}$ and $G_i$ is a Hermitian operator. Following an analogous argument to that in Sec.~\ref{app:corol} we find that 
$ \Var_{\chi_{lm}^{SD}}[J_k] \leq  || G_k^2 ||_{\infty}$.  Defining $G_{\rm max} = \text{max} \{ ||G_k^2||_{\infty} \}$, we have that
\begin{equation}\label{eq:full_gen_variance_ineq}
\begin{aligned}
     \Var_{\mathbb{V}}[\partial_{\theta_k} C_{\rm gen}(\vec{\theta}, V)] &\leq w_{\text{max}}^2 \left[\frac{2 X_{\max}}{2^{2n}-1}+\frac{2 Y_{\max}}{2^n(2^{2n}-1)}\right]G_{\rm max} \, 
\end{aligned}
\end{equation}
where we use $\sum_l \tilde{p}_l = 1$.
If $ \Tr[H_l^S H_m^S  ] \in \mathcal{O}( 2^n)$ and $\Tr[H_l^S] \in \mathcal{O}(2^n)$, it follows that $X_{\rm max} \in \mathcal{O}( 2^n)$ and $Y_{\rm min} \in \mathcal{O}( 2^{2n})$. Therefore, additionally assuming that $w_{l} = w_{ijk} \in \mathcal{O}(1)$ and $||G_k^2||_{\infty} \in \mathcal{O}(1)$, we find that
\begin{equation}
\begin{aligned}
       \Var_{\mathbb{V}}[\partial_{\theta_k} C_{\rm gen}] \in \mathcal{O}( 2^{-n}) \, 
\end{aligned}
\end{equation}
as claimed. 





\end{proof}

\section{Expected Scaling for Typical Ans\"{a}tze}\label{ap:ScalingAnsatzes}
We argue that under some practical circumstances, the cost function gradient can be suppressed further. First note that
the summation in the variance involves a large number ($\sim 2^n$) of states $\{|\psi_i\rangle\}$ which in general are not orthogonal to each other. The typical values of the overlaps $|\langle\psi_i|\psi_j\rangle |^2$ scale as $2^{-n}$. A typical unitary ansatz also appears as random, which allows us to estimate the typical values of each term in $\Var_{\chi_{ij}^S}[J_k]$, i.e.,
\begin{equation}
\begin{aligned}
\int dU\  \left(\text{Tr}[\chi_{ij}^SJ_k]\right)^2
       =& \int dU\  \text{Tr}\left[ U^{\dag} J_k U |\psi_j\rangle\langle\psi_j| U^\dag J_k U |\psi_i\rangle\langle\psi_i| \right]\\
        =& \frac{\left(\text{Tr}[J_k]\right)^2 \ + \text{Tr}[J_k^2]}{2^{2n}-1} 
 - \frac{\left(\text{Tr}[J_k]\right)^2 \ + \text{Tr}[J_k^2] |\langle\psi_i|\psi_j\rangle|^2}{2^n(2^{2n}-1)} \sim \frac{1}{2^n}
\end{aligned}
\end{equation}
and
\begin{equation}
\begin{aligned}
\int dU\  \text{Tr}[\chi_{ij}^SJ_k^2] \text{Tr}[\chi_{ij}^S]
       =\frac{ |\langle\psi_j|\psi_i\rangle|^2\text{Tr} [J_k^2]}{2^n} \sim \frac{1}{2^n} \ \text{for}\ i\ne j  .
\end{aligned}
\end{equation}
Therefore, a $\sim 2^{-2n}$ scaling may commonly be observed in the variance of the cost function gradient.

\section{Cost function to learn an unknown unitary}\label{ap:Numericsdetails}

\begin{figure}[t]
\centering
\includegraphics[width=\linewidth]{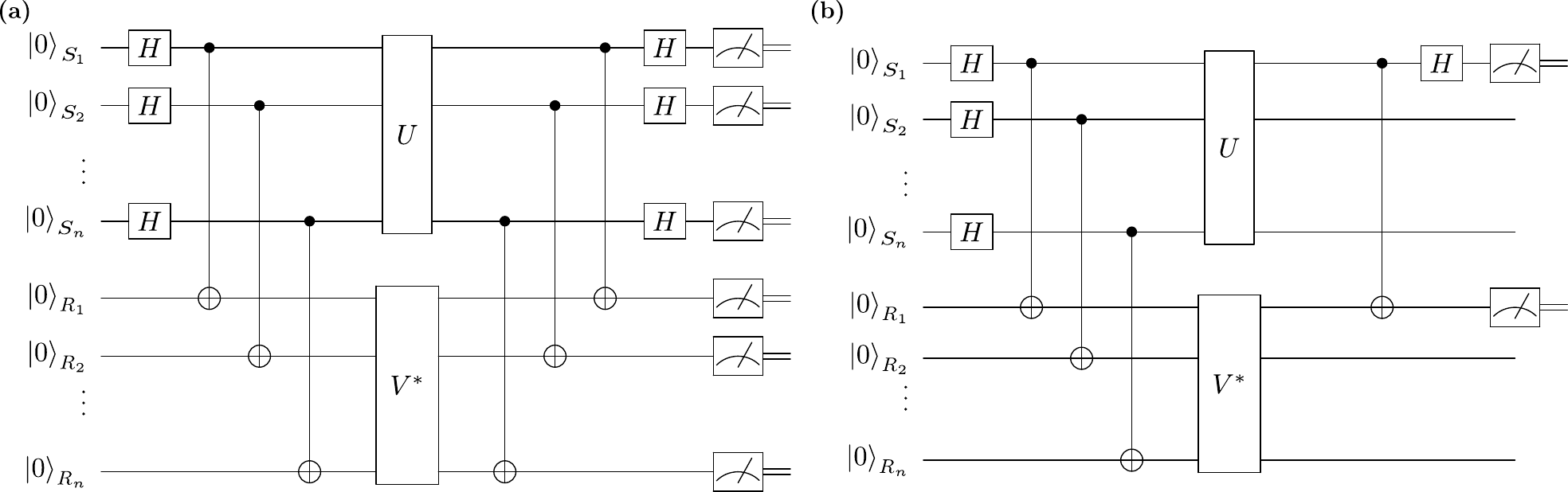}
\caption{\label{fig:Circuits} (a). The Hilbert-Schmidt Test. The probability to measure the all zero state at the end of the circuit is given by $\langle \Phi_+ | (U V^\dagger \otimes \I_R ) |\Phi_+ \rangle \langle \Phi_+ | ( V U^\dagger \otimes \I_R) |\Phi_+ \rangle   = \frac{1}{2^{2n}} | \Tr[U V^\dagger ] |^2$. (b) The Local Hilbert-Schmidt Test. The probability to measure zeros across qubits $S_j$ and $R_j$ is equal to $\langle \Phi_+ | (U V^\dagger \otimes \I_R ) \I_{\overline{S}_j}\otimes |\phi^+ \rangle\langle \phi^+|_{S_jR_j} \otimes \I_{\overline{R}_j}  ( V U^\dagger \otimes \I_R) |\Phi_+ \rangle $. Note, in both circuits we have used the ricochet property $ (X^\dagger \otimes \I )\ket{\phi_+} = (\I \otimes X^* )\ket{\phi_+} $ for any operator $X$. (Figure adapted from Ref.~\cite{khatri2019quantum}).}
\end{figure}

A natural choice in cost function to learn an unknown unitary $V$ can be formulated in terms of the Hilbert-Schmidt inner product between $V$ and a trainable unitary $U$ as follows
\begin{equation}
    C_{\rm HST}(U, V) = 1 - \frac{1}{2^n}|\Tr[U V^\dagger]|^2  \, .
\end{equation}
In Ref.~\cite{khatri2019quantum} this cost was demonstrated to have the following desirable properties. 
\begin{enumerate}
    \item It is faithful, vanishing iff $U$ and $V$ agree up to a global phase $\phi$, that is iff $U = V \exp(-i \phi)$.
    \item  It is operationally meaningful since it can be related to the average gate fidelity between $U$ and $V$.
    \item It can be efficiently computed on a quantum computer. 
\end{enumerate}
To see the latter, note that $C_{\rm HST}(U, V)$ can be written as
\begin{equation}
    C_{\rm HST}(U, V) = 1 - \langle \Phi_+ | (U V^\dagger \otimes \I_R ) |\Phi_+ \rangle \langle \Phi_+ | ( V U^\dagger \otimes \I_R) |\Phi_+ \rangle 
\end{equation}
where $\ket{\Phi_+}$ is a maximally entangled state across two registers each containing $n$ qubits. Thus the cost takes the form of Eq.~\eqref{eq:MoreGencost} in the main text with $\rho_{\rm HST} = |\Phi_+ \rangle \langle \Phi_+ | $ and $H_{\rm HST} = \I - |\Phi_+ \rangle \langle \Phi_+ | $. That is, the computer is prepared in a maximally entangled state across the two registers, the first register is evolved under $U V^\dagger$, and then finally a Bell state measurement is implemented across the two registers. The circuit to perform this protocol, and thereby measure $C_{\rm HST}(U, V)$, is known as the Hilbert-Schmidt Test and is shown in Fig.~\ref{fig:Circuits}(a). We further note that the ricochet property $ (X^\dagger \otimes \I )\ket{\Phi_+} = (\I \otimes X^* )\ket{\Phi_+}$ for any operator $X$ implies that $(U V^\dagger \otimes \I )\ket{\Phi_+} = (U \otimes V^* )\ket{\Phi_+}$. This is used in Fig.~\ref{fig:Circuits}(a) to apply $U$ and $V$ in parallel and thereby reduce the depth of the cost function circuit.  

While the form of $C_{\rm HST}(U, V)$ is intuitive, for larger systems it exhibits barren plateaus even for short depth ans\"{a}tze~\cite{CerezoBP2020}. For this reason, in the numerical implementations performed in this paper we use a `local' variant of the Hilbert-Schmidt Test which avoids this problem. The local cost, $C_{\rm LHST}$, can be written in the form of Eq.~\eqref{eq:MoreGencost} with $\rho_{\rm LHST} = |\Phi_+ \rangle \langle \Phi_+ | $ (similarly to $C_{\rm HST}$) but now with $H$ composed of a sum of projectors onto the local Bell states $\ket{\phi_+}_{S_jR_j}$ on the qubits $S_j$ and $R_j$. Specifically, let us consider two $n$-qubit registers $S$ and $R$ and let $S_j$ ($R_j$) represent the $j_{\rm th}$ qubit from the $S$ ($R$) register. The $C_{\rm LHST}$ cost is of the form $C_{\rm LHST} = \frac{1}{n} \sum_{j= 1}^{n} C_{\rm LHST}^{(j)}$ where 
\begin{equation}
   C_{\rm LHST}^{(j)} = \langle \Phi_+ | (U V^\dagger \otimes \I_R ) H^{(j)}_{\rm LHST}  ( V U^\dagger \otimes \I_R) |\Phi_+ \rangle
\end{equation}
and 
\begin{equation}
    H_{\rm LHST}^{(j)} = \I_{SR} -  \I_{\overline{S}_j}\otimes |\phi^+ \rangle\langle \phi^+|_{S_jR_j} \otimes \I_{\overline{R}_j} \, .
\end{equation} 
Here $\overline{S}_j$ denotes the set of all qubits in $S$ except for $S_j$, and similarly for $\overline{R}_j$. The circuit used to measure $C_{\rm LHST}$ is shown in Fig.~\ref{fig:Circuits}(b). 

In Ref.~\cite{khatri2019quantum} it was proven that $C_{\mbox {\tiny LHST}}$ and $C_{\mbox {\tiny HST}}$ are related as 
\begin{equation}\label{eq:LHSTvsHSTbound}
    C_{\mbox {\tiny LHST}}(U,V) \leq C_{\mbox {\tiny HST}}(U,V) \leq n \, C_{\mbox {\tiny LHST}}(U,V) \, .
\end{equation}
Consequently, $C_{\mbox {\tiny LHST}}$ inherits $C_{\mbox {\tiny HST}}$'s desirable properties. Specifically, $C_{\mbox {\tiny LHST}}$ vanishes iff $C_{\mbox {\tiny HST}}$ vanishes, and hence $C_{\mbox {\tiny LHST}}$ is faithful. Furthermore, $C_{\mbox {\tiny LHST}}$ can be shown to bound the average gate fidelity.

\section{Average and Variance of Gradient for Approximate Designs}\label{Ap:approx}

In this section we discuss the average and variance of the gradient, when the target ensemble is not an ideal but approximate 2-design.

Let $\mu$ be a distribution of a unitary ensemble, and $\Delta_\mu$ a 2-fold channel
\begin{equation}
    \Delta_\mu (\rho) = \int_\mu\ dU  U^{\otimes 2} \rho (U^\dag)^{\otimes 2}.
\end{equation}
Here we use a strong definition for an approximate unitary 2-design~\cite{Brandao2016-vu}, namely, a unitary ensemble with distribution $\mu$ is an $\epsilon$-approximate 2-design iff
\begin{equation}\label{app:ordering}
    (1-\epsilon)\Delta_{Haar} \preceq \Delta_\mu \preceq (1+\epsilon)\Delta_{Haar},
\end{equation}
where $\preceq$ is the semi-definite ordering, i.e., channel $\mathcal{A} \preceq \mathcal{B}$ iff $\mathcal{B}-\mathcal{A}$ is completely positive.

The second order integral we are interested in can be expressed as
\begin{equation}
   \text{Tr} \int_\mu dU \ A U B U^\dag D U E U^\dag = \text{Tr}\left[ A\otimes E \Delta_\mu(B\otimes D) W_p \right],
\end{equation}
where $W_p$ is the permutation operator on the tensor product Hilbert space. Denote $F_U = \text{Tr} A U B U^\dag D U E U^\dag$. When $A$, $B$, $D$ and $E$ are all positive operators, as a result of the semi-definite ordering (\ref{app:ordering}), 
\begin{equation}
(1-\epsilon) \int_{Haar} dU F_U \le \int_\mu dU F_U \le (1+\epsilon) \int_{Haar} dU F_U.
\end{equation}
This relation carries over to non-positive operators, by expanding with positive operators as a basis set in the operator space. By choosing $D=E=\I$, the first order integral involved in the average of the gradient is also covered by the above relation.

Hence, for an $\epsilon$-approximate 2-design, the average gradient is also zero, and the variance is equivalent to the case of ideal 2-designs up to a multiplicative factor $1+\epsilon$.

\clearpage

\end{document}